\documentclass[reprint,onecolumn,amsmath,amssymb,aps,
]{revtex4}

\usepackage{epsfig,bm,epsf,graphics}
\usepackage{graphicx}
\usepackage{dcolumn}
\usepackage{bm}

\begin{document}
\preprint{APS/123-QED}

\title{Monte Carlo Study of  Agent-Based Blume-Capel  Model for Political Depolarization}

\author{Hung T. Diep$^{1}$\footnote{diep@cyu.ftr},  Miron Kaufman$^{2}$\footnote{m.kaufman@csuohio.edu},  Sanda Kaufman$^{3}$\footnote{s.kaufman@csuohio.edu}}
\affiliation{%
$^1$  Laboratoire de Physique Th\'eorique et Mod\'elisation,
CY Cergy Paris Universit\'e, CNRS, UMR 8089\\
2, Avenue Adolphe Chauvin, 95302 Cergy-Pontoise Cedex, France.\\
$^2$  Department of Physics, Cleveland State University, USA.\\
$^3$ Levin School of Urban Affairs, Cleveland State University, USA.}



\date{\today}


\begin{abstract}

 In this paper, using Monte Carlo simulations we show that the Blume-Capel model gives rise to the social depolarization. This model borrowed from statistical physics uses the continuous Ising  spin varying from -1 to 1 passing by zero to express the political stance of an individual going from ultra-left (-1) to ultra-right (+1).  The particularity of the Blume-Capel model is the existence of a $D$-term which favors the state of spin zero which is a neutral stance. We consider the political system of the USA where voters affiliate with two political groups: Democrats or Republicans, or are independent. Each group is composed of a large number of interacting members of  the same stance. We represent the general political ambiance (or degree of social turmoil) with a temperature $T$ similar to thermal agitation in statistical physics. When three groups interact with each other, their stances can get closer or further from each other, depending on the nature of their inter-group interactions. We study the dynamics of such variations as functions of the value of the $D$-term of each group. We show that the polarization decreases with incresasing $D$. We outline the important role of $T$ in these dynamics. These MC results are in excellent agreement with the mean-field treatment of the same model.
\end{abstract}

\maketitle


%
%
%
%

\section{Introduction}

Social polarization has been investigated by numerous scholars using many methods [1-6]. In democracies, each person is free to choose his/her stance with respect to a social issue such as politics, immigration, race, economics, ... according to his/her preferences. People with the similar beliefs regarding specific sets of social issues form groups that can be very large, such as political parties [2]. Policies proposed by the left are very different of those proposed by the right. Polarization derives from sharp differences between individuals affiliated with the "left" orientation and those with a "right" orientation [3,4]. Depending on the social context and events, the political polarization at any moment can be very sharp: people belonging to different political parties may not agree on almost any subject. The cases of France and of the USA in recent times are striking. For example, since 2022 in France, no political party holds a majority in the parliament. Therefore, in the absence of compromise between at least two parties, the majority required to pass a law does not materialize. Instead, most of the time the government uses article 49.3 of the Constitution to pass laws without a parliament vote [6,7]. In the USA, the political polarization began to rise in the 70's [8-13]. We see this in periodic polls [11]. Other european democracies have seen the same tendency of strong political polarization [16,17].

In general, each party proposed policies to solve societal problems ranging from government aid to the needy, to race, immigration, national security, and the environment [6]. However, in democracies, protracted conflict over solutions to societal problems leads to more problems: political polarization has serious deleterious societal [1,16,17,18] and economic [3] consequences. One of these is that people gradually lose the ability to work together, to compromise, make, and implement deals. In time, this can lead to societal breakdown [17,19]. 

One notable problem caused by polarization is that individuals increasingly tend to believe in information which agrees with, or justifies their political perspective [17,20,21]. In time, the information sets from which they draw support diverge to the point where polarized groups hold entirely contradictory images of the shared reality.
Therefore, strong polarization prevents constructive debates. This may lead to political instability [17]--the governing party changes often, leading to collective and individual uncertainty. Changing the governing party at each round of elections prevents implementation of long-term programs which are necessary in realms such as economics and education. Since polities are complex systems [22] within which interactions change with time, according to [23] and [24] empirical studies do not suffice to help us understand political polarization dynamics. We also need theoretical modeling to help explore the conditions under which specific events can happen. Agent-based modeling has great potential in this regard ([25,26]). Modeling can help prepare information that might be useful in reducing the impacts of polarization [27,28].  

Together with the recognition of increasing polarization, there has been a rise in the number of investigations of its causes and dynamics. Sociophysics--namely applying physics tools to the study of social phenomena--has been a very effective approach in this respect. It can handle complexity in various domains, including politics, and provide insights complementing those gleaned from other disciplines [28-33]. Sociophysics  has already been used in studies of polarization (e.g., [10,20,34-37]). Network models can be used to explore polarization trends and to find avenues for intervention, generating qualitative anticipatory scenarios which can be queried (e.g., [38-43]).  For instance, using network models we have anticipated election outcomes in the US and in Bosnia-Hercegovina [28,42]; and we have examined various outcomes of labor-management contract negotiations in France [43]. As Ref. [44] has argued, anticipatory scenarios are useful in supporting the development of robust strategies of action in the face of the high levels of uncertainty characterizing complex systems.

Within Western democracies political polarization is on the rise, undermining collective decision making abiity (e.g., [17,45,46]). We have examined polarization dynamics in the USA between Democratic- and Republican-affiliated individuals, using an agent-based model borrowed from statistical physics using mean-field theory and Monte Carlo (MC) simulations  [47,48].

We note that very recently Galam [49] has studied the political polarization using his model of opiniion dynamics which consists in supposing there are several categories of indiividuals in a communinity, each with a probability: the contrarians, the floaters and the stubborn agents. He found several scenarii of polarization  depending on the case and the probability:  from the unanimity to the rigidity passing by the coexistence. His  method is probabilistic while ours  uses the spin model with microscopic interaction between individuals of the same group at time $t$ and interactiion of individuals with the average stance of the other groups at the earlier time $t-1$.  We believe however that qualitatively we should find the same kinds of polarization if we modify our model to match with his assumptions.

We note also that there exist several other physical models that can be mapped into the social language to describe social phenomena.  Let us mention that the polarization between charged particles [50] can be seen as a social polarization, or the mean-field approach treating the  separation of ionic liquids described via the Cahn-Hilliard term in a regular solution can be also seen as a collective polarization [51].

Once strong polarizaton is present, we need to search for ways to reduce it, namely to depolarize the society. We have used the Blume-Capel model from statistical physics [52,53] to study depolarization using the mean-field approximation [54]. As descibed below, this model has a term (called $D$-term) which favors the neutral position in each individual, which may collectively reduce polarization. 

In  [54], agents’ interactions had an infinite range (mean-field), meaning that each individual interacting with each of the others in a political system with three groups. As in [48], here we extend our work by assuming, instead, that individuals interact only with their “neighbors”. We explore the insights to be gained with the short-range interactions assuming a Bravais lattice, which may be more realistic in terms of how individuals communicate and try to persuade others to their political stance. Moreover, this kind of short-range interaction matches a “massively parallel” approach proposed by [55] as a practical means of reducing polarization. Our model may help assess the extent to which the massively parallel approach can be effective in reducing polarization. Note that agent-based modeling has been used to study attitude change in societies [56]. 

In section \ref{MM} we describe the initial Blume-Capel model [54] and its counterpart short-range model we use for Monte Carlo simulations in this paper. In section \ref{RD} we present the simulation results and discuss their meaning in terms of scenarios of depolarization trends. We conclude in section \ref{conclu} with a summary.

\section{Model and Method}\label{MM}
\subsection{Model}

Let us briefly explain the origin of depolarization in our recent paper ([54]), where we have used agent-based modeling to extend a sociophysics 2-group network model of conflict dynamics [38] to three political groups in the US: Democrats (group 1), Republicans (group 2), and Independents (group 3).

To describe the political stance of an individual in a society, we use a spin $S$ model, where the attitude $S$ ranges between -1 (extreme left) and 1 (extreme right). An individual’s stance can take any value on the continuum within this range. Such a model is called a ”continuous Ising model” ([57], [58]), as opposed to the discrete Ising model, where $S$ could only take two values, -1 and 1. We use this continuous $S$ model here in the Blume-Capel model described below. To compare with the mean-field approximation [54], we  use in this paper MC simulations with short-range interactions between individuals, with real-time fluctuations.

Each individual in group $i$ ($i=1,2,3)$ has a stance compatible with the group’s attitude $S_i$ regarding a specific issue under debate– economics, social issues, defense, etc.—or (here) a package of such issues (in the [1] and [6] sense). The individual stances have values between -1 and +1, where -1 corresponds to the democrats/progressive/left position ($i=1$), while +1 corresponds to the republicans/conservative/right position $(i=2)$. Individuals thus align with the group whose average stance is compatible and closest to their own [1]. 

Inside groups 1 and 2, individuals are homophilic [5]: they tend to prefer to communicate with each other, rather than with individuals from a different group. We denote $J_i$ the link between members of group $i$. It quantifies the cohesiveness of group $i$. Through $J_i$, members inside each group attempt to persuade each other to their own stance, effectively diminishing intra-group differences and causing stances to converge. 

Individuals in each group also keep an eye on the other groups’ average attitudes, which in turn influence their own, either nudging the group average to a more extreme value or to a more moderate one. These inter-group interactions are described by parameters $K_{ij}$. For group 1, the inter-group interaction terms, -$K_{12}S_1<S_2>$ and -$K_{13}S_1<S_3>$, represent the influence of the mean stances of groups 2 and 3, $<S_2>$ and $<S_3> $ respectively, on an individual in group 1. The inter-group interactions $K_{12}$ and $K_{21}$ are not necessarily equal. At times, members of one group may feel cooperative toward the other, who might not reciprocate. Therefore, in general, $K_{ij}\neq K_{ji}$  because of human agency. While physics phenomena obey Newton’s third law, the magnitudes of human action and reaction do not have to be equal. Rather, the effect of group $i$ on group $j$ can be different in magnitude and sign from the effect group $j$ has on group $i$.  Hence our model is not described by a single Hamiltonian and its dynamics is not the Glauber dynamics (our spin is not Ising +/-1 but is continuous). 
A temperature $T$, reflecting contextual factors, acts on each individual indepently, the way the thermodynamic temperature acts on particles.

The intra-group cohesion parameters $J$ and the inter-group influence parameters $K$ affect the average group attitudes in time. For instance, according to the recent Gallup polls [45], in early 2023 40\% of adults declared themselves independent—with zero internal cohesion $J_3$, since they are not organized or formally linked, like Democrats or Republicans, but rather a bin for the non-affiliated. However, in February 2023 all but 7\% of them leaned either Democrats or Republicans, at least partly in response to persuasion efforts by the other two groups. 

We also use a magnetic field $h_i$ to represents the effect of group i’s leadership on group's members. When $h_i > 0$, group i’s mean stance is nudged toward positive values; when $h_i < 0$ the mean stance is nudged to negative values.

The model’s Hamiltonian of group $i$ is inspired from the time-independent Blume-Capel model. It is given by:

\begin{equation}\label{H0}
H_i (t)=-J_i\sum_{m,n} S_i (m,t)\cdot S_i(n,t)+D_i\sum_iS_i^2(m,t)- h_i\sum_m S_i(m,t)
\end{equation}
where $i$ indexes group $i$, and $S_i(m,t)$ is the stance of an individual $m$ in group $i$ at time $t$. The sum is performed over the nearest neighbors (NN) $m$ and $n$ belonging to  group $i$. Note that for group 3 (Independents), the first sum is zero because $J_3=0$. Also, a group at $t$ interacts with the average stances of the other groups at $t-1$. 

Note the that the positive sign of the $D$ term favors the small value of $S_i(m,t)$ when $D_i$ is positive. Smaller $S_i$ causes smaller polarization. In orther words, positive $D$ is at the origin of the {\it depolarization}, as will be shown below.

When the three groups interact, the Hamiltonians of each group is as follows:
\begin{eqnarray}
{\cal H}_1(t) &=& H_1(t)-K_{12}\sum_m S_1(m,t)<S_2(t-1)>\nonumber\\
&&-K_{13}\sum_m S_1(m,t)<S_3(t-1)>\label{H1}
\end{eqnarray} 
\begin{eqnarray}
{\cal H}_2(t) &=& H_2(t)  -K_{21}\sum_m S_2(m,t)<S_1(t-1)> \nonumber\\
&&-K_{23}\sum_m S_2(m,t)<S_3(t-1)>\label{H2}
\end{eqnarray}
\begin{eqnarray}
{\cal H}_3(t)&=& H_3(t)  -K_{31}\sum_m S_3(m,t)<S_1(t-1)> \nonumber \\
&&-K_{32}\sum_m S_3(m,t)<S_2(t-1)>\label{H3}
\end{eqnarray}

The model has the following parameters: three $J_i$, three $D_i$ for the three groups’ respective internal cohesiveness and "anisotropy", three $K_{ij}$ and three $K_{ji}$ describing the inter-group interactions (note that $K_{ij}$ and $K_{ji}$ are  not  necessarily symmetric),  and three $h_i$ to describe leadership effects, if any. The $J$ and $K$ parameters can be selected qualitatively, as we have done below, using publicly available poll data (see also [6]; [39]; [33]). 

	%
	%


We have solved this model using the mean-field approximation [54]. We define a polarization measure $P$ as the distance between the mean stances of groups 1 and 2 at a given time $t$:

\begin{equation}\label{eq-polar}
P = (<S_2>-<S_1>)/2
\end{equation}

such that $-1 \leq P \leq 1$. $<S_i>$ is the average individual stance of group $i$ calculated at a time $t$. When $P = 0$, there is no polarization. It occurs when groups 1 and 2 have equal average stances $<S_1>=<S_2>$. When $P = 1$, polarization is extreme (also called hyperpolarization (e.g., Burgess et al [17]). This can occur when Democrats’ stance $<S_1>= -1$ (most progressive/left) and Republicans’ stance $<S_2>= 1$ (most conservative/right).$P$ can be negative if  $<S_1>$ and $<S_2>$ change their signs (it can be in politics).

Here, we use a similar model, but with short-range intra-group interactions and perform MC simulations. 

Before proceeding to the simulation method, let us discuss the role of the “political” temperature $T$  introduced below, and borrowed from statistical physics. The temperature in statistical physics represents thermal agitations of the particles (spins, for example). Thus $T$ acts as a disordering factor: at low $T$, particles stay in the lowest energy state (or very close to it), while at high $T$, they vigorously change their state in an independent manner, causing disorder in the system in spite of the inter-particle interaction which favors order. The well-known example is the ferromagnets: spins in ferromagnets are parallel at low $T$ but become disordered at high $T$. In the context of political groups considered here, $T$ represents the political ambiance of the society. When an election is not imminent, or the society is calm, $T$ is low. Each group is relatively stable, with no significant effect of inter-party interaction. Close to an election or during politically fraught times with important issues at stake  such as strained economies or international tensions, intra-party cohesiveness may wane, due to the fluctuation of individual stances of its members, equivalent to high “political” temperature $T$. Then each group might attempt to take advantage of the weakened cohesiveness of the other groups to enhance its influence in the competition. As we shall see below, $T$ plays an important role in outcomes of politic contests.

\subsection{Simulation Method}\label{simul}

We take the case of the US political system: there are  three groups: Group 1 (Democrats), Group 2 (Republicans) and Group 3 (Independents). We assume Group 1 to have a stronger cohesiveness (largest $J$), and to be governing. Group 2 has weaker cohesiveness and is in opposition. Group 3 is composed of individuals having no unified political framework, and no formal intra-group communication links ($J_3=0$). The Independents are often attracted to the stance of the opposition party, they play a contrarian role  (see [59-61] for other examples of contrarian used in a model). 

\begin{figure}[ht]
\centering
\includegraphics[width=4cm,angle=0]{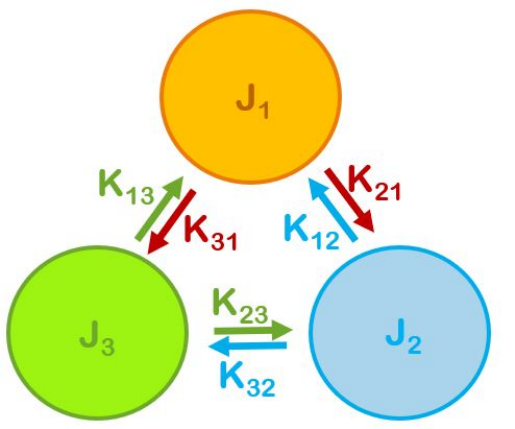}
\caption{Interaction parameters. 
Note that $J_3$, intra-group interaction among the Independents, is zero. See text for comments.}
\label{notation}
\end{figure}

For the MC simulations, we represent each of the three groups with a triangular lattice of size $N\times N$, where each site is occupied by a member. Each member interacts with its six nearest neighbors (NN) at time $t$, and considers the average stances of the other groups calculated at $t$ – 1 (a realistic lag). The choice of this lattice allows for a maximum number of NN in 2D. Of course, we can use a 3D lattice to have more NN  such as a FCC or a HCP with 12 NN. However we believe it will not give new phenomena.  For each group, we use the periodic boundary conditions to reduce the size effects. In general, we take the size of 100 × 100 lattice sites for each group. See Figure \ref{notation} for the interaction parameters described in the previous subsection. Note that the notion of NN interactions in politics does not necessarily mean that people are geometrically close to each other. Rather, it refers to the number of people generally in contact with an individual.

To simulate the three groups' interactions at a given $T$, we carry out the  simulation as follows: for a group we generate initial individual stances -1 (Democrats) for group 1,  1 for group 2 (Replublicans), and 0 (Independents) for group 3. We can also use the initial stances all equal to zero for three groups.  We use next the Metropolis algorithm to find the collective state of each group at the time $t$, taking into account the average stances of the other groups at $t-1$,  as described in Eqs.  (\ref{H1})-(\ref{H3}). We follow the evolution of each group with time $t$ (MC time). 

The Metropolis algorithm used for  updating the individual stance of a member is as follows: at a time $t$, we calculate the interaction energy $E_{old}$ of a member with its NN and with an effective field resulting from the two other groups at time $t – 1$. We make a trial change of its state by choosing a random stance between -1 and 1. We calculate the member’s trial new energy $E_{new}$. If $E_{new} < E_{old}$, the trial state is accepted.  If $E_{new} > E_{old}$, it is accepted with
the probability $\exp[-(E_ {new} - E_{old})/(k_BT)]$. We repeat this updating procedure for all individuals in each of the three groups.
Note that the Metropolis algorithm obeys the detailed balance only when the system is at  equilibrium, namely when there is a probability conservation: state A to state B has the same probability with that from B to A. Our purpose is to study the time dependence of the polarization, so there is no such probability conservation. Note that there are several popular dynamics such as Glauber dynamics and Kawasaki dynamics, but to our knowledge all of them have been devised for discrete spins, not for continuous spins used in this paper.  The advantage of the Metropolis algorithm is that it does not depend on the nature of spin, it can be used for any kind of spin such as continuous spins  used here, XY spins or Heisenberg spins.

\section{Results and Discussion}\label{RD}

As seen, our model has 9 principal interaction parameters  $J_i(i=1,2,3)$ and $K_{ij}(i\neq j, i=1,2,3,j=1,2,3$ in addition  to $D_i (i=1,2,3)$ and $h_i (i=1,2,3)$. However, in applications the choice of the parameters is limited. As discussed in [47,48,54], this choice is guided by polls  [45,46] and by political common attitudes of the people:
to produce anticipatory scenarios of polarization, we made the following assumptions:\\
-The Democrats (group 1) are more cohesive than Republicans (group 2), i.e., $J_1 > J_2$;\\
-Independents (group 3) have no cohesion ($J_3 = 0$) because they have no structure
or means of identifying with each other, do not communicate, and do not recruit;
therefore, they exert no influence on the other two groups and, as such,$ K_{13} = K_{23} = 0$;\\
-Independents tend to be contrarian to the party in power (here, group 1), thus $K_{31} < 0$,
and are not influenced by the opposition party, thus $K_{32} = 0$.

With respect to parameter value selection, guided by media and professional, frequent polling reports, we assigned parameter values such that they qualitatively mimic general polls results [45,46]. To enable a comparison of MC results with those obtained with the MFT model [54], we selected the same values for parameters $J$ and $K$, as follows: \\

\begin{eqnarray}
\mbox{- Intra-group interactions:}&& J_1 = 5,  J_2 = 3,\nonumber\\
&&   J_3 = 0\label{para1}\\
\mbox{- Inter-group interactions:}&&K_{12} = -4, K_{21} = -5, \nonumber\\
K_{13 }= 0,  K_{31} = -3,&& K_{23} = 0,  K_{32} = 0.\label{para2}
\end{eqnarray}

\noindent     - Depolarization parameters $D_i$: we will choose several cases presented below.\\

Note that a negative $K_{ij}$ indicates hostility (or resistance) of group $i$ toward group $j$, while a positive sign indicates attraction or potential agreement between two groups.  A variation of the above values keeping their signs will not alter qualitatively the results shown in the following.

Each group is thermalized at temperature $T$  in interaction with the other groups.  
We have calculated the following quantities:\\
\\

- Cohesive energy per individual  $E_i(T)= <{\cal H}_i(T)> /N^2$ where $<{\cal H}_i(T)>$ is the thermal average at $T$ given by
\begin{equation}\label{HT}
<{\cal H}_i(T)>=\sum_{t=t1}^{t2}{\cal H}_i (t)/(t2-t1)
\end{equation}
where $t1$ is the starting averaging time and $t2$ the averaging end time. The total cohesive energy $E(T)\sum_i E_i(T)$ is also calculated. \\
\\

- Stance of each group (sublattice magnetization) as a function of $T$:
\begin{equation}
M_i(T)= < S_i (T)> = \sum_{t=t1}^{t2} \sum_n S_i (n,t)/(t2-t1)/N^2                              
\end{equation}
where $n$ belongs to group $i$.  Within the assumption of the parameters given above one has $M_1<0$, $M_2>0$. In the absence of $D$, $M_3>0$.
We define the strength of group $i$ by $Q_i(T)=|<S_i>|,$\\
\\

- Susceptibility or fluctuations of the stance of group $i$ at $T$:
\begin{equation}
\chi (T) =[<M_i(T)^2>-<M_i (T)>^2]/(k_BT)
\end{equation}

\subsubsection{The case $D_1=D_2=0$ and $D_3=5$}
\begin{figure}[ht]
\centering
\includegraphics[width=4cm,angle=0]{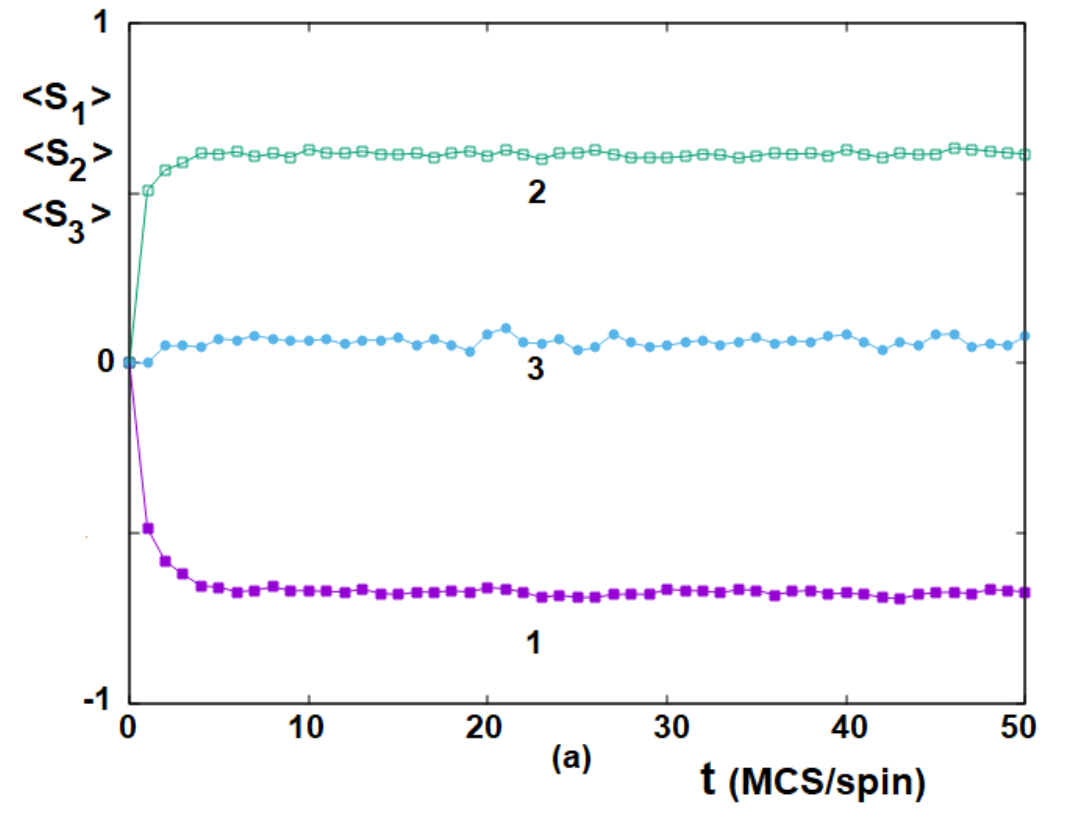}
\includegraphics[width=4cm,angle=0]{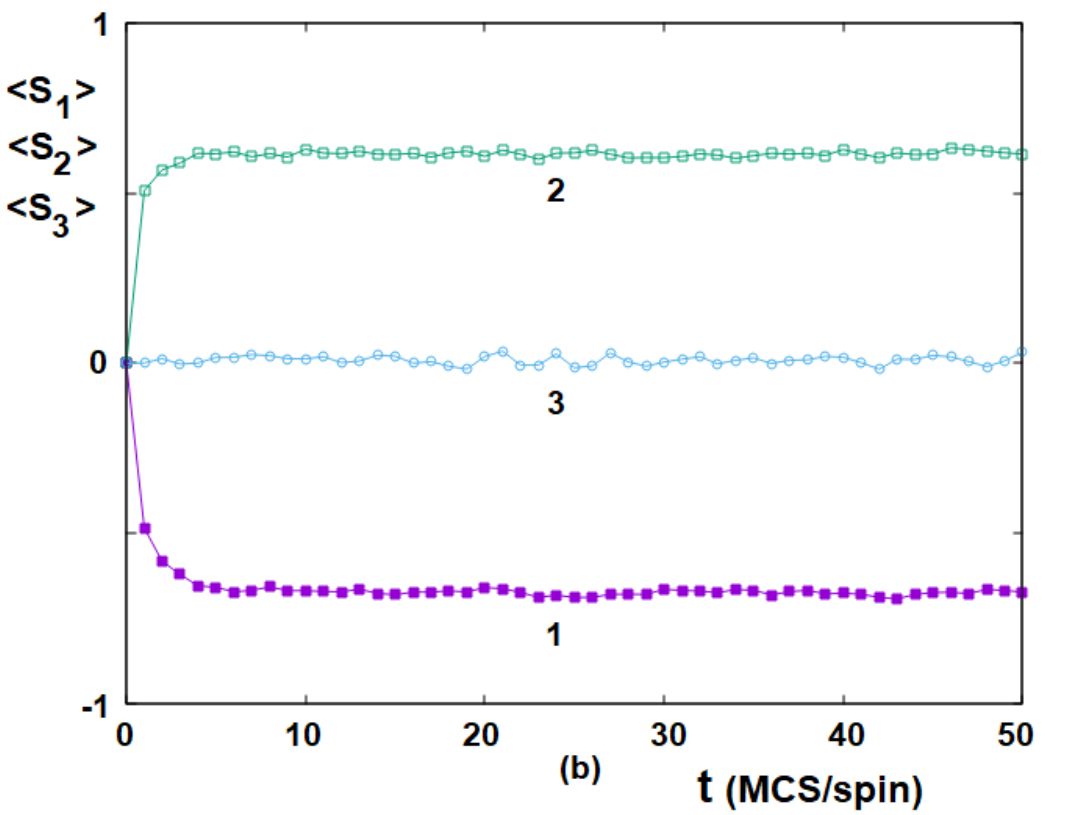}
\caption{Stances of the three groups with $D_1=D_2=0$ and (a) $D_3=0$; (b) $D_3=5$ as function with MC time. Other interaction parameters are given in Eqs. (\ref{para1})-(\ref{para2}). Curves 1, 2 and 3 correspond respectively to $<S_1>$, $<S_2>$ and $<S_3>$.
Note that $J_3$, intra-group interaction among the Independents, is zero. See text for comments.}
\label{fig2}
\end{figure}

 With $K_{31}=-3$ (Group 3 is contrarian to Group 1), if $D_3=0$ the stance of Group 3 should be positive, though small but visible  as seen in Fig. \ref{fig2}a. But when $D_3=5$ as considered here, the stance of Group 3 is neutralized by $D_3$. It is zero as shown in Fig. \ref{fig2}b.

\subsubsection{The case $D_1=D_2=3$ and $D_3=5$} 

\begin{figure}[ht]
\centering
\includegraphics[width=4cm,angle=0]{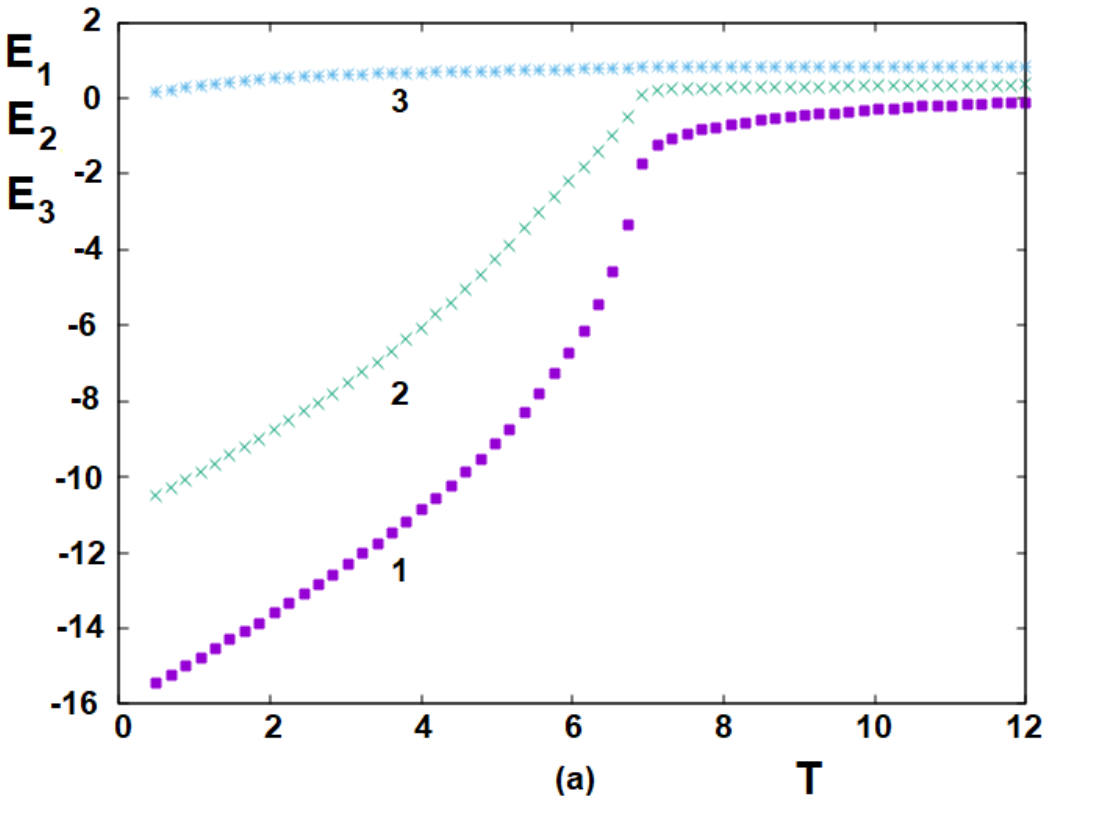}
\includegraphics[width=4cm,angle=0]{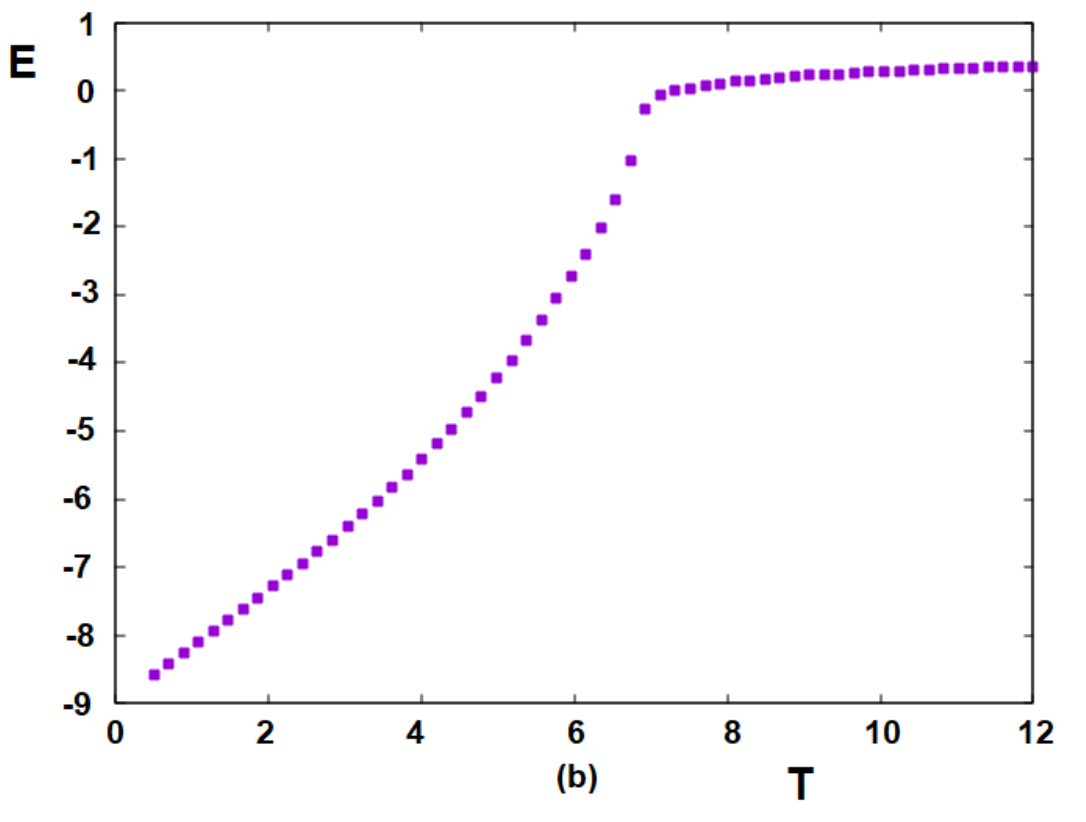}
\includegraphics[width=4cm,angle=0]{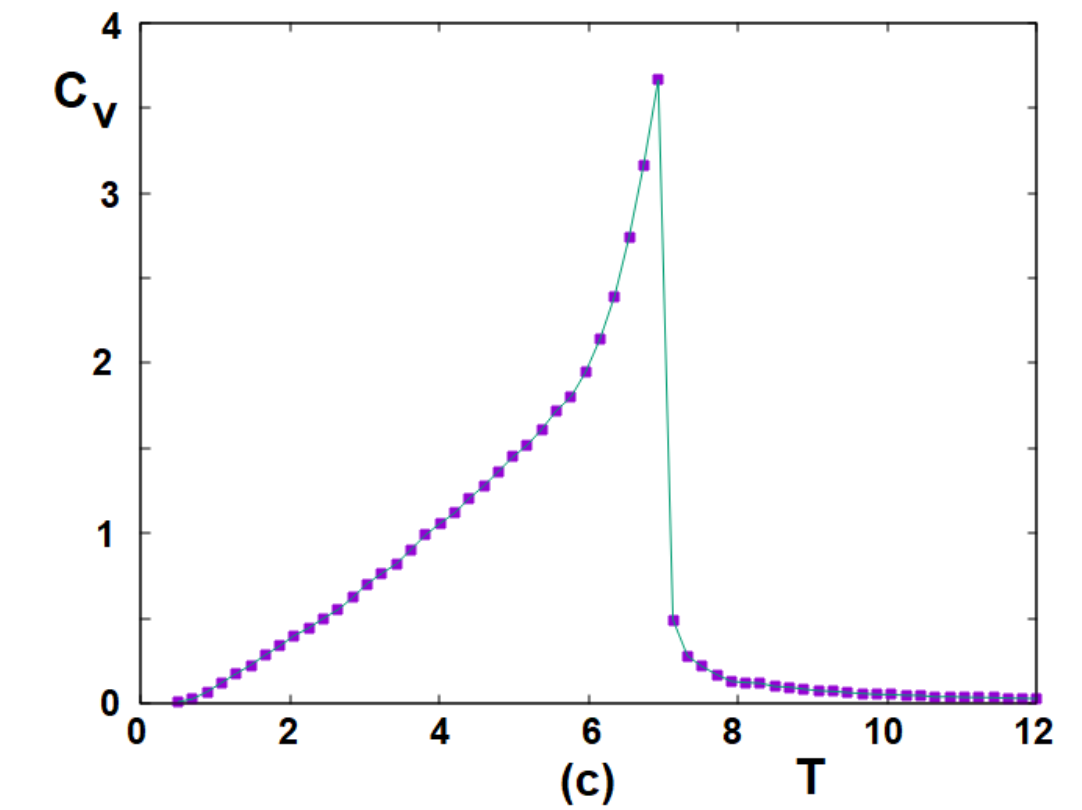}
\caption{(a) Internal energy of three groups, curves 1, 2 and 3 correspond respectively to Group 1, Group 2 and Group 3; (b) Total energy per individual ; (c) Total  specific heat per individual,  as functions of political temperature $T$. See text for comments.}
\label{fig3}
\end{figure}

\begin{figure}[ht]
\centering
\includegraphics[width=4cm,angle=0]{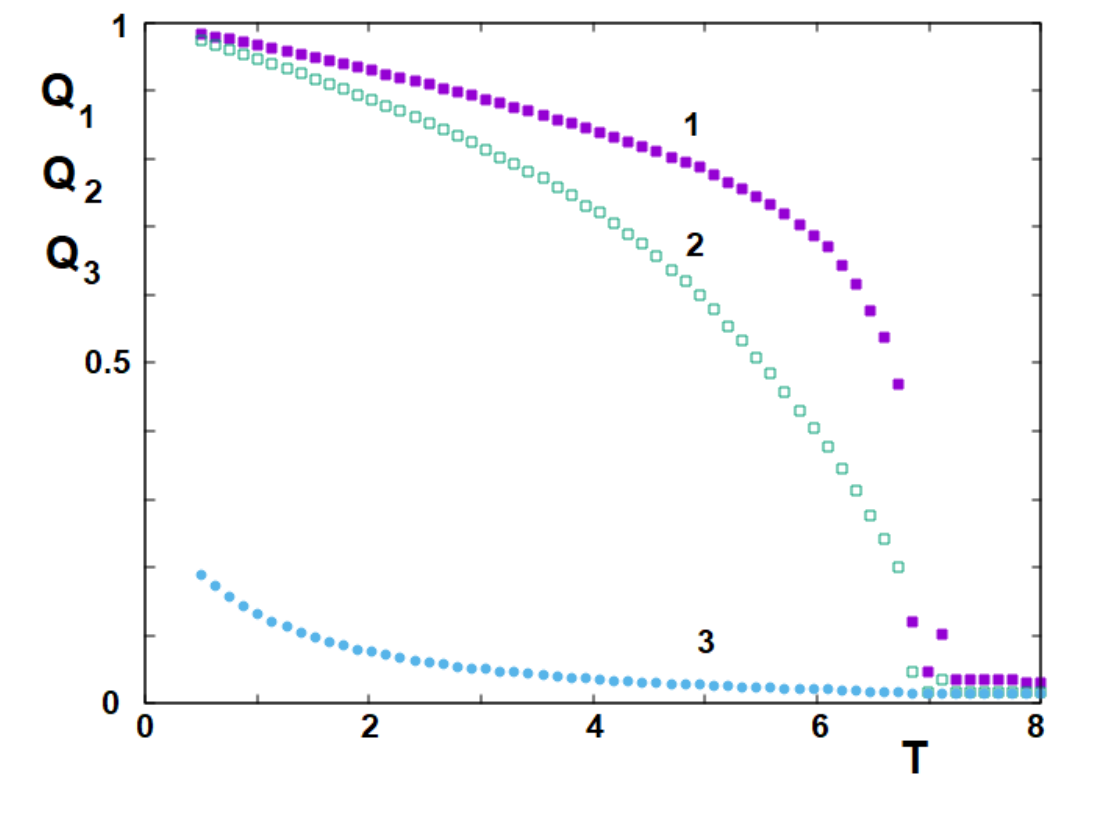}
\caption{Strengths $Q$ of three groups  as functions of political temperature $T$. Curves 1, 2 and 3 correspond respectively to Group 1, Group 2 and Group 3.}
\label{fig4}
\end{figure}

\begin{figure}[ht]
\centering
\includegraphics[width=4cm,angle=0]{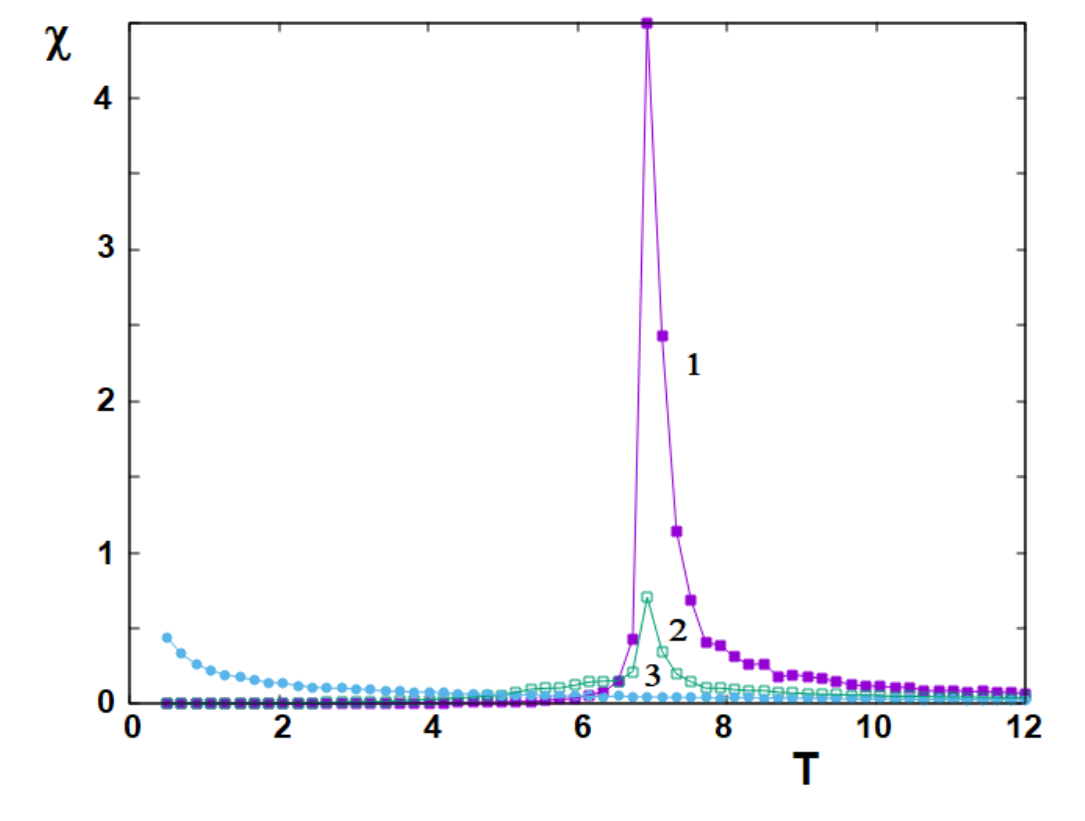}
\caption{Fluctuations of the three groups' stances as functions of political temperature $T$. Curves 1, 2 and 3 correspond respectively to Group 1, Group 2 and Group 3.}
\label{fig5}
\end{figure}

 The equilibrium cohesive energy of each group is shown in Figure \ref{fig3}a,  the total cohesive energy of three groups  is displayed in Figure \ref{fig3}b as a function of temperature $T$. Since the three groups interact with each orther, there is only a transition temperature $T_C \simeq 6.93 $ where all of them become disordered (the energy changes its curvature  at this point). The specific heat per individual is shown in Figure \ref{fig3}c where the peak temperature corresponds to $T_C$.  In terms of sociophysics, above $T_C$ there is no cohesiveness between individuals.  The society is in a turmoil state.
Note that the energy of Group 3 is due to its interaction with group 1. This is confirmed in Figure \ref{fig4}, showing the absolute values of stances $Q_1=| < S_1 > |$, $Q_2=| < S_2 > |$ and $Q_3=| < S_3 > |$: we see that $Q$ is higher for larger $J$, and all drop to zero (i.e., no cohesiveness) at $T_C$. In statistical physics, $T_C$ is called transition temperatures [58] above which the systems become disordered. At and close to $T_C$, the stances of the groups strongly fluctuate as shown in Fig. \ref{fig5}. These fluctuations of the order parameter in statistical physics correspond to the so-called susceptibilities which are the fluctuations of $M_i $, namely $(< M_i^2 > - < M_i >^2)/(k_BT)$.

%

\begin{figure}[ht]
\centering
\includegraphics[width=4cm,angle=0]{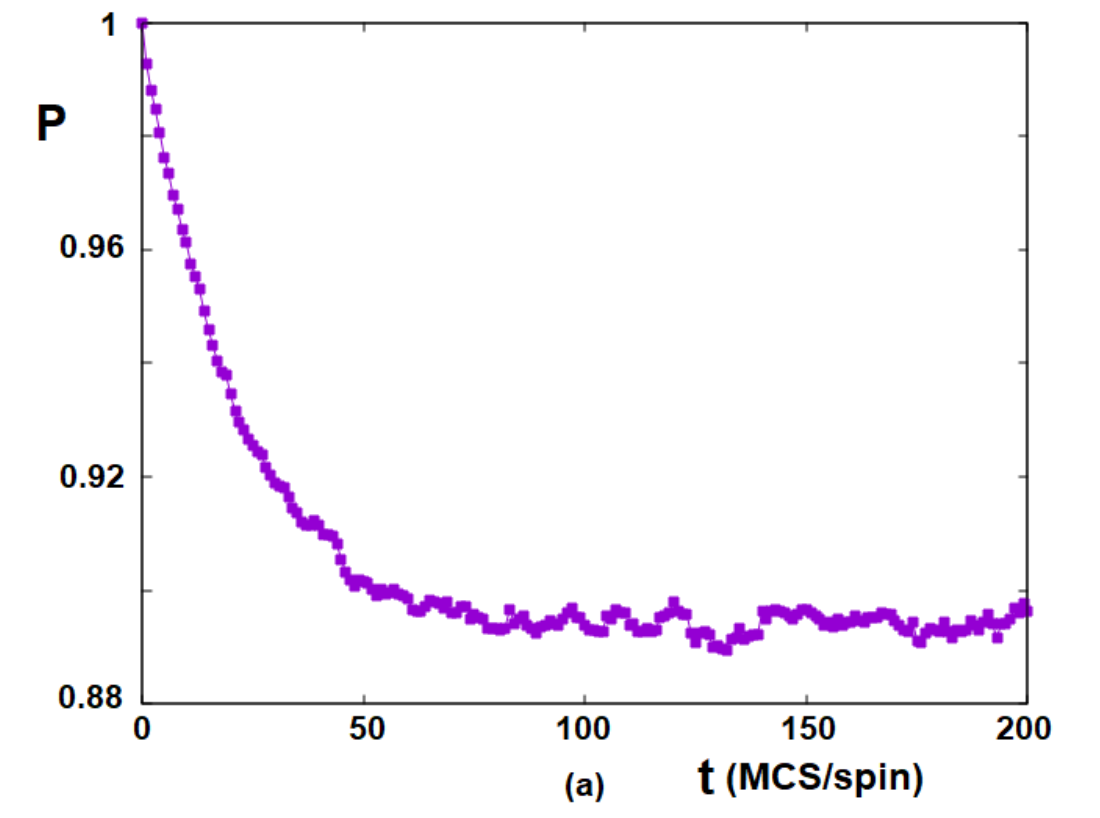}
\includegraphics[width=4cm,angle=0]{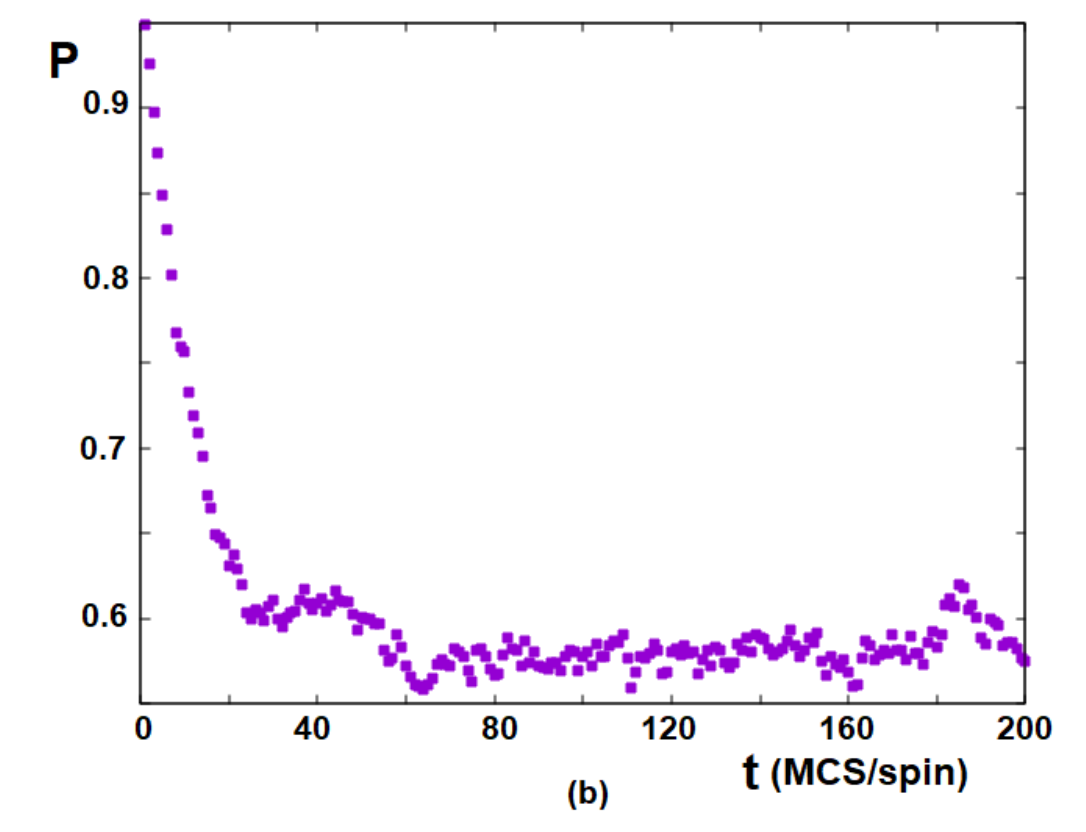}
\caption{Political polarization as a function of time $t$ at (a) $T=2.254\simeq T_C/3$; (b) $T=5.762$ close to $T_C$. See text for comments. }
\label{fig6}
\end{figure}

Let us show now the polarization with evolving time $t$. Starting from the highest state of polarization $P=1$, namely $<S_1>=-1$, $<S_2>=+1$, we see that $P$ diminishes to  0.89 (see Fig. \ref {fig6}a) calculated at a low temperature $T=2.254$, and to  0.58 at $T=5.762$. These values are much lower than the case where there is no depolarization $D$ (see [48]).   The role of $T$ is very important since $P$ strongly depends on $T$.

\subsubsection{The case of higher $D_1$, $D_2$ and $D_3$} 

\begin{figure}[ht]
\centering
\includegraphics[width=4cm,angle=0]{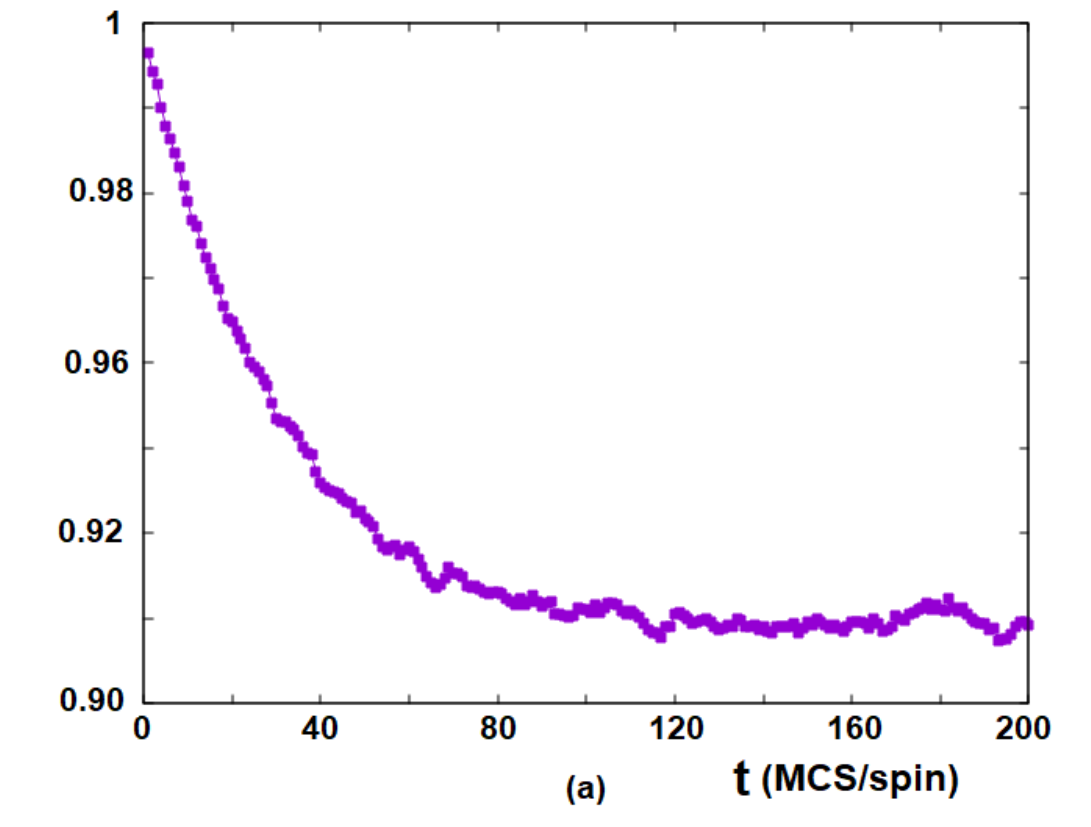}
\includegraphics[width=4cm,angle=0]{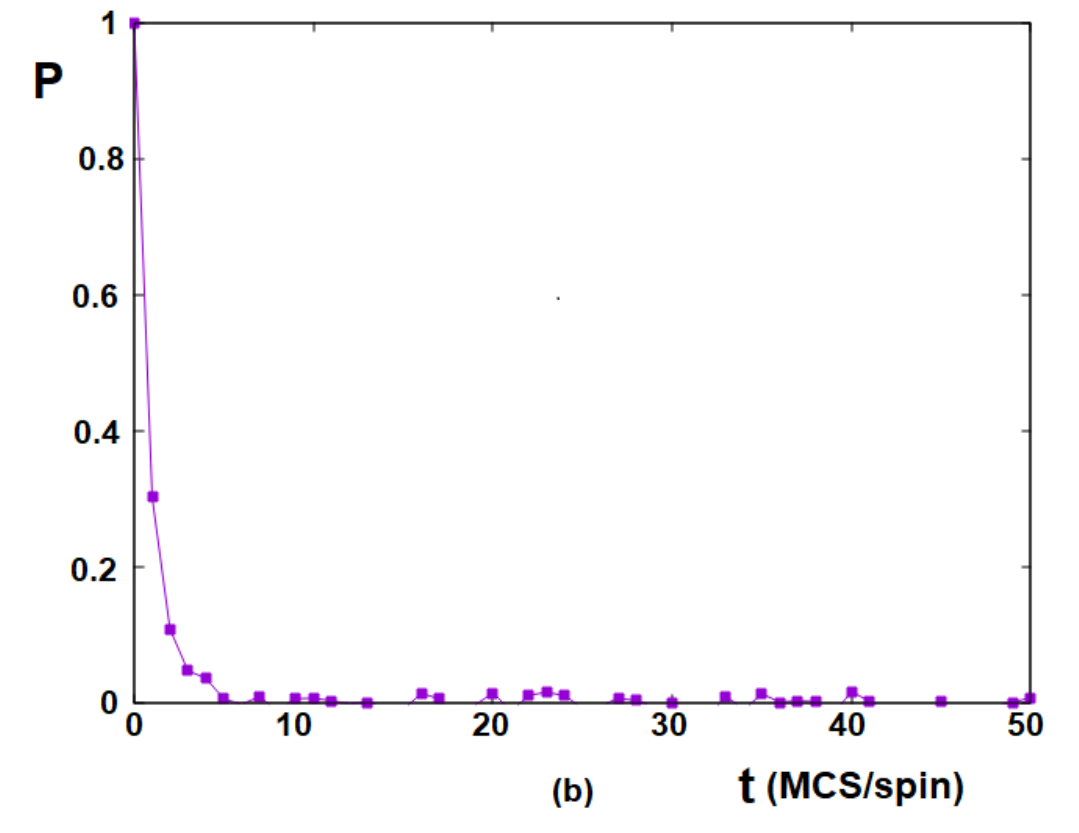}
\caption{Political polarization as a function of time $t$ at (a) $D_1=D_2=D_3=5$ at  $T=1.90<< T_C$; (b) for $D_1=D_2=D_3=10$ at $T=3.2$ slightly above $T_C$. See text for comments. }
\label{fig7}
\end{figure}

Let us take the cases of stronger depolarization  $D_1=D_2=D_3=5$ ($T_C=5.84$) and $D_1=D_2=D_3=10$  ($T_C=3.29$)  . We show in Fig. \ref{fig7} the polarization versus time $t$ .

\begin{figure}[ht]
\centering
\includegraphics[width=4cm,angle=0]{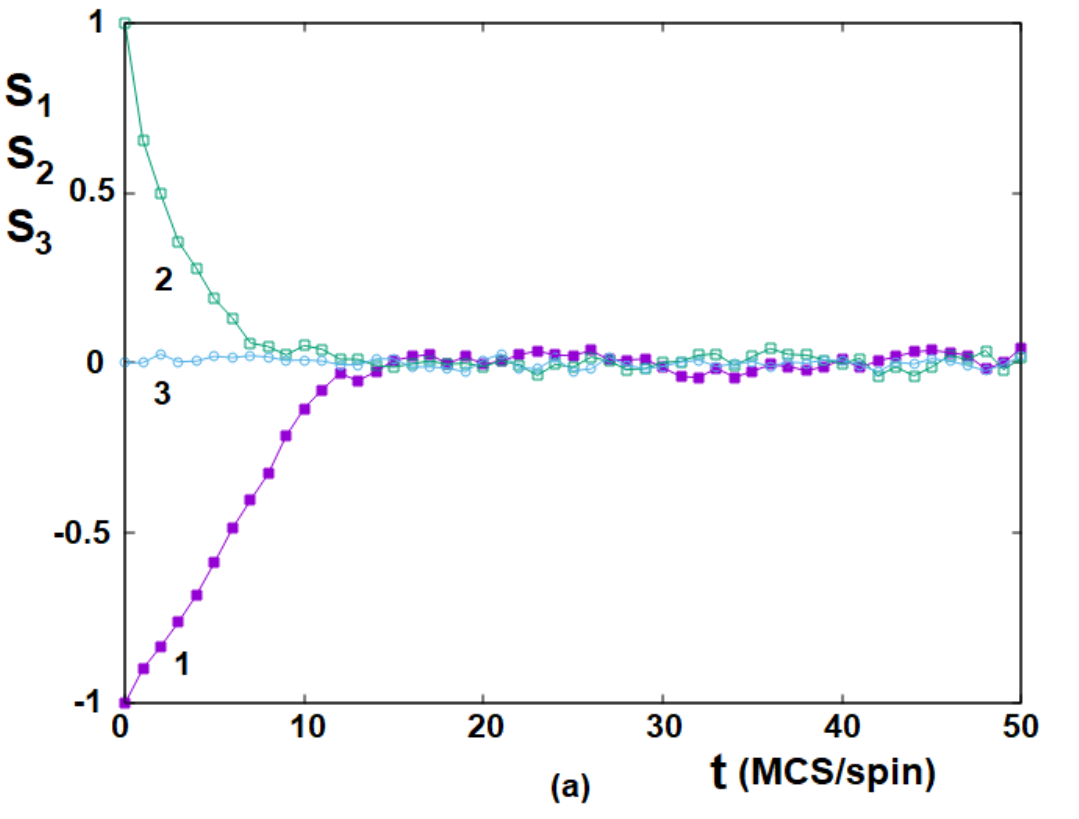}
\includegraphics[width=4cm,angle=0]{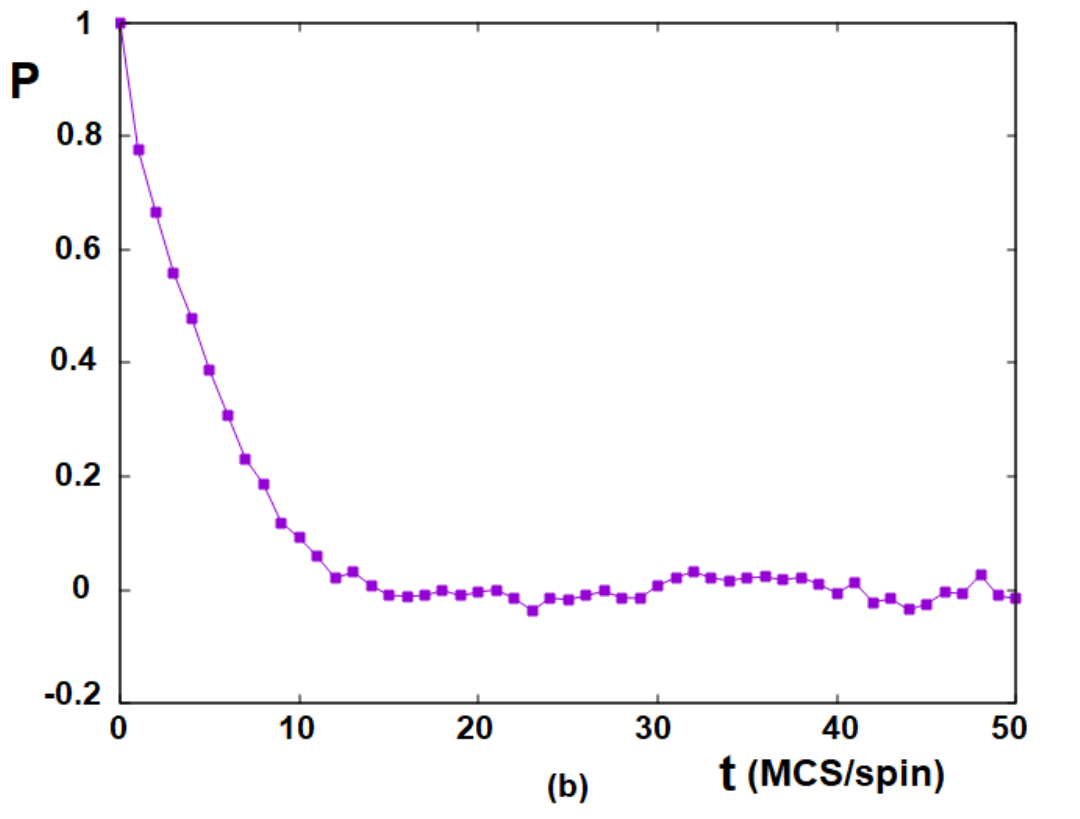}
\caption{The case $D_1=D_2=D_3=5$: (a) Stances  of three groups versus $t$ at a political temperature $T$ higher than $T_C$, curves 1, 2 and 3 correspond respectively to $<S_1>$, $<S_2>$ and $<S_3>$. (b) Polarization.}.
\label{fig8}
\end{figure}

If we go close to $T_C$ in all cases, we have a strong damping of the group stances and of the polarization versus $t$.  We show in Fig. \ref{fig8} one example in the case  $D_1=D_2=D_3=5$ at $T=6$ slightly above $T_C$.

\subsubsection{Strong oscillation of the polarization}

\begin{figure}[ht]
\centering
\includegraphics[width=4cm,angle=0]{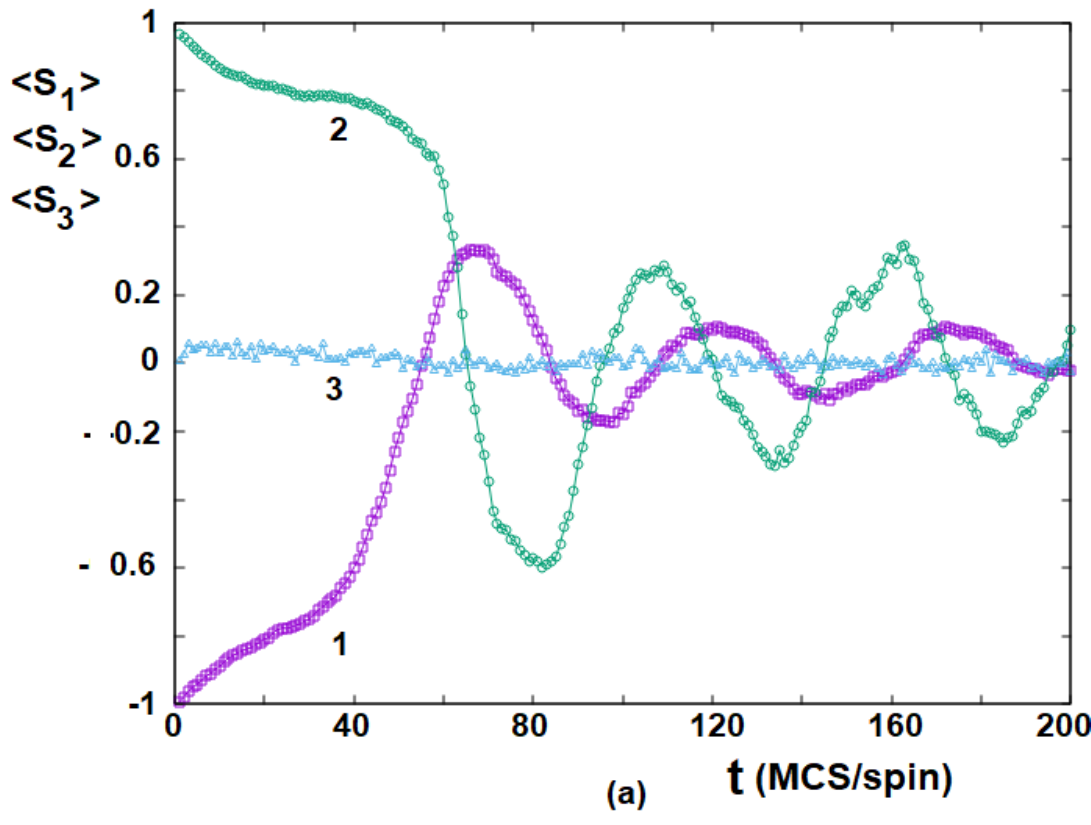}
\includegraphics[width=4cm,angle=0]{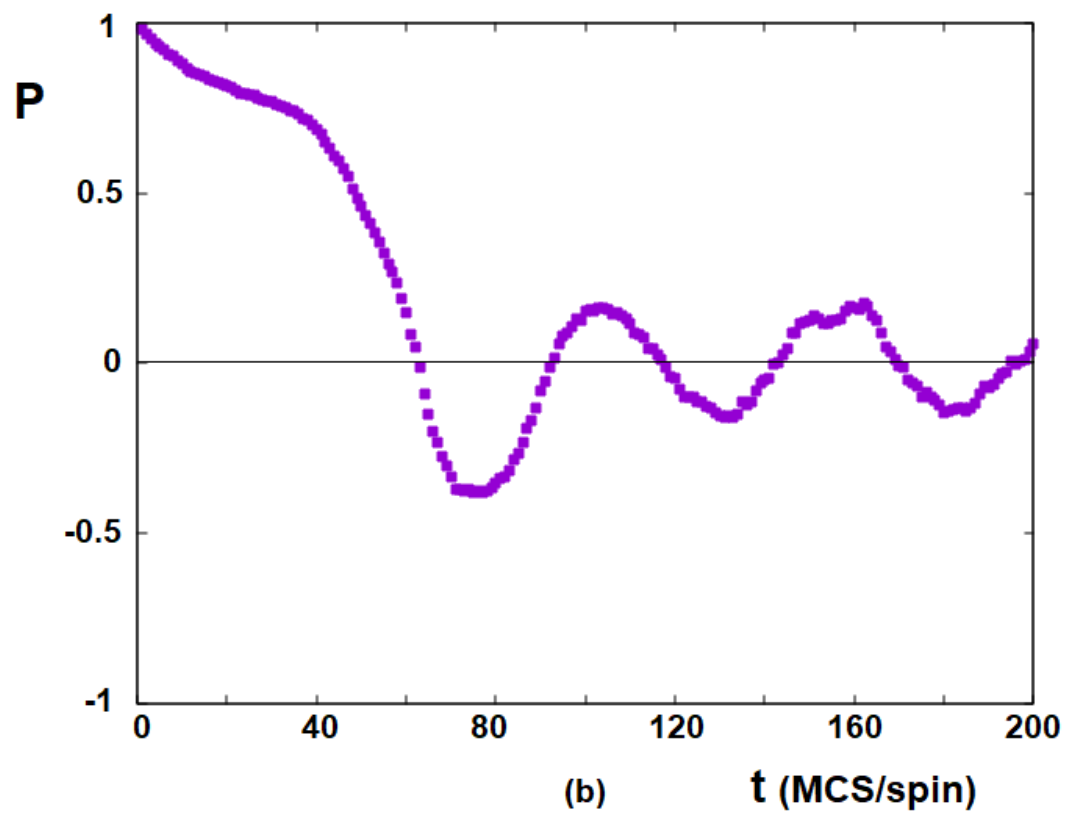}
\caption{(a) Political stances as a function of time $t$, curves 1, 2 and 3 correspond respectively to $<S_1>$, $<S_2>$ and $<S_3>$;   (b) Polarization versus $t$. The parameters are $D_1=D_2=3$, $D_3=5$ with   $K_{21}=6$ and $K_{12}=-5$.  See text for comments. }
\label{fig9}
\end{figure}

One of the interesting cases is when  the interactions between Group 1 and Group 2 have the opposite signs. Let us take the case $D_1=D_2=3$ and $D_3=5$ and $K_{21}=6$ and $K_{12}=-5$. We show in  Fig. \ref{fig9} the oscillating stances of the three groups and the oscillation of the polarization. This phenomenon occurs in a large temperature region below $T_C$. The amplitude of the oscillation depends on the values of $K_{21}$ and $K_{12}$.  

The attraction of the Republicans to the Democrats (positive $K_{21}$) is the origin of the oscillation of $P$ in time.  We recall that $P$ measures the distance between the political stances.  

At this stage, it is worth to emphasize that oscillation phenomena are found in several systems due to the tendency of the matter to self-organize:  the phenomenon was first discovered by Boris Belousov in 1951, while he was trying to find the non-organic analog to the Krebs cycle. When mixing potassium bromate, cerium(IV) sulfate, malonic acid, and citric acid in dilute sulfuric acid, he observed that the ratio of concentration of the cerium(IV) and cerium(III) ions oscillated, causing the colour of the solution to oscillate between a yellow solution and a colorless solution. This is due to the change of cerium(IV) ions  by malonic acid into cerium(III) ions, which are then oxidized back to cerium(IV) ions by bromate(V) ions. The reader is referred to Ref. [62] for a review. In our model, the oscillation occurs when a party is  attracted so far from its equilibrium by the other party, its intra-cohesive energy gets it back.  It is some kind of a restoring force analog to the oscillatory motion of a pendulum.  

To conclude this section, we have performed MC simulations on the same statistical physics model as the one where we used the mean-field approximation (see [54]). Despite the fact that the mean-field model neglects fluctuations while MC simulations take into account space and time fluctuations, the two methods yield qualitatively the same patterns of political polarization. 
  
\section{Conclusion}\label{conclu}

We have proposed the Blume-Capel model to explore, using MC simulations, whether depolarization is possible between Democrats and Republicans in the USA as a function of time. 
The same model has been studied by mean-field theory by our group [54].  
We have considered three groups with initial different political stances: Democrats, Republicans, and Independents. An individual within any of these  groups interacts with a limited number of people sharing the same political viewpoint. At any time, individuals also consider the average stance of other groups in the previous time period, causing them to either become firmer or soften their stance. Although the model represents the political structure in the USA, it can be adapted to other three-group dynamics.

We find that MC results for short-range intra-group interactions agree well with those obtained by the mean-field which assumed long-range interactions.  
The Blume-Capel model assumes that each individual is represented by a continuous Ising spin taking its values from -1 (left, liberal orientation) to +1 (right, conservative orientation). What is important in our model is the fact that there is the $D$ term in the Hamiltonian, which can soften the opposite stances, which we called depolarization.The model shows that the depolarization stems from the effect of the $D$ term on individual orientations. It may be more efficient/durable than collective measures taken by party leaders.  
 
The MC simulation results show that polarization depends on $D$ and on the nature of the inter-group interactions. It may advantage the party in opposition and help it win an election. It may also give rise to an oscillation of the polarization (whose sign changes in time). Therefore, the outcome of an election depends on the moment in time when it occurs.
It is interesting to note that the MC and mean-field models yield qualitatively very similar results with the same depolarization dynamics. Both approaches can be used to generate scenarios that include various interventions to reduce polarization. The MC near-neighbor approach lends itself to generating scenarios for another kind of intervention proposed by [55] and [17], who have called it “massively parallel.” It consists of independent individuals and groups operating locally to reach out and initiate dialogues with people holding opposite stances, thereby reducing the current acute homophily. Such initiatives are already taking place around the USA (see [55]).

The two versions of the Blume-Capel models--mean-field, with infinite-range interactions and  Monte Carlo method, with near-neighbor interactions, show ways to depolarize society and find practical actions which might correspond to various $D$ values. Some on-going empirical studies and opinion polls will provide data which we plan to include qualitatively in our models to refine the scenarios we can generate, and make them relevant to spceific contexts such as the United States and France.

In conclusion, we have illustrated how agent-based models from statistical physics which contain sufficient ingredients can concisely describe complex situations in social sciences which may appear intractable (for example, in terms of number of variables and data) when studied with traditional, non-dynamic methods.  



\end{document}